\documentclass[traditabstract]{aa}
\usepackage{txfonts}
\usepackage{epsfig}
\usepackage{graphicx}
\usepackage{longtable}
\usepackage{booktabs}
\usepackage{natbib}
\usepackage{url}
\usepackage{twoopt}
\usepackage[breaklinks=true]{hyperref}
\usepackage{amstext}

\begin{document}  

\title{Collisional effects in  the blue wing of Lyman-$\alpha$}

\author{F.~ Spiegelman\inst{1} 
  \and
  N.~F.~Allard   \inst{2,3}
  \and
  J. F. Kielkopf  \inst{4}
  }

   \institute{Laboratoire de Physique et Chimie Quantique, Universit\'e de      
           Toulouse (UPS) and CNRS, 118 route de Narbonne, 
           F-31400 Toulouse,   France \\
     \and
     GEPI, Observatoire de Paris,  Universit\'e PSL, 
     UMR 8111, CNRS,
     61, Avenue de l’Observatoire, F-75014 Paris, France\\
          \email{nicole.allard@obspm.fr}
     \and
Sorbonne Universit\'e, CNRS, UMR7095, Institut d'Astrophysique
de Paris, 98bis Boulevard Arago, PARIS, France\\
\and
Department of Physics and Astronomy, 
    University of Louisville, Louisville, Kentucky 40292 USA \\
}

\date{fevrier2021; }

\abstract{Spectral observations below Lyman-$\alpha$ are now obtained
  with the {\it Cosmic Origin Spectrograph} ({\it COS}) on the {\it
    Hubble Space Telescope} ({\it HST}). It is therefore necessary to
  provide an accurate treatment of  the blue wing of the
  Lyman-$\alpha$ line that enables correct calculations of radiative
  transport in DA and DBA white dwarf stars.  On the theoretical
  front, we very recently developed very accurate H-He
  potential energies for the hydrogen $1s$, $2s$, and $2p$ states.
  Nevertheless, an uncertainty remained about the asymptotic
  correlation of the $\Sigma$ states and the electronic
  dipole transition moments.  A similar difficulty occurred in our
  first calculations
  for the resonance broadening of hydrogen perturbed by collisions with neutral
  H atoms. 
   The aim of this paper is twofold.
   First, we  clarify the question of the asymptotic  correlation of
   the $\Sigma$ states, and we show  that relativistic contributions, even  very tiny, may need to be accounted for a
   correct long-range and asymptotic description of the states because of   the specific $2s§2p$ Coulomb degeneracy in hydrogen. 
   This effect of relativistic corrections,  inducing  small splitting of
   the $2s$ and $2p$ states of H, is shown to be  important for the 
   $\Sigma$-$\Sigma$ transition dipole moments in H-He and is also discussed
   in H-H.
   Second, we use existent (H-H) and newly determined (H-He) accurate
   potentials and properties  to 
   provide a  theoretical investigation  of the  collisional effects on
   the blue wing of the Lyman-$\alpha$ line of
  H perturbed by He and H.
  We  study the relative contributions in the blue wing of the
  H and He atoms according to their relative densities.
  We  finally  achieve a comparison with recent {\it COS}
   observations and propose an assignment for a feature centered at
  1190~\AA\/.

}
 

\maketitle
\titlerunning{Collisional effects in  the near blue wing of Lyman-$\alpha$}
\authorrunning{Spiegelman \& Allard}

\section{Introduction}
\label{sec:introduction}

In helium-dominated white dwarfs, the discrepancy in
the hydrogen abundance between Balmer-$\alpha$ from the optical data and Lyman-$\alpha$
from the ultraviolet (UV) data is strong. In addition, the Lyman-$\alpha$ line profile is
asymmetric \citep[see][and references therein]{xu2017}.
The existence of a quasi-molecular line satellite  is crucial
      for understanding  this asymmetrical shape of the
      Lyman-$\alpha$ line observed with
the {\it Cosmic Origin Spectrograph} ({\it COS})
      (see Fig.~1 in \citet{allard2020}).
This absorption feature has been  predicted by detailed 
collisional broadening profiles by \citet{allard2009d}.
These authors made an exhaustive study of the red wing of the
Lyman-$\alpha$  line perturbed  by H-He collisions.
They considered high He densities met in cool DZ
white dwarfs and  examined the validity range of the one-perturber
approximation that is widely used to calculate  the line wings.
H-He potentials were theoretically determined by several authors, namely \citet{theodora1984}, \citet{theodora1987}, \citet{sarpal1991}, \citet{lo2006},
 \citet{belyaev2015}, and  \citet{allard2020}.  \citet{allard2009d} used  the potentials and dipole moments of
\citet{theodora1984} and \citet{theodora1987}, but were limited
by a lack of  accuracy of the molecular potential of the $C\Sigma$ state.
They noticed an unexpected  well of about 150~cm$^{-1}$  at $R \sim 8$~\AA\/
that is related to the choice of basis states.
Significant progress in the description of the H-He potential energies
has been achieved in a recent work \citep{allard2020} using  extensive
basis sets and multi-reference configuration interaction (MRCI)
calculations (\citep{knowles92} and \citep{molpro2015})
 to determine accurate ab initio potential energy curves.
 Nevertheless, because of the specific  degeneracy of the hydrogen levels in
 the Coulomb model,
the adiabatic correlation of the $A$ and $C$ states  to either 2$s$
(dipole forbidden from the ground state) or 2$p$ (allowed)
states is  not  fully characterized using the  Coulomb Hamiltonian only.
 Relativistic
effects that are  smaller than 1 cm$^{-1}$ for hydrogen are  responsible for
lifting the strict degeneracy of the hydrogen atomic levels in the Coulomb model~\citep{kramida2010}. This level splitting
is crucial for establishing the adiabatic  correlation of the molecular states  toward the  asymptotic levels, 
and thus the specific asymptotic behavior
of the dipole transition moments from the ground state.  
Thus, one aspect of the present paper is to  redetermine and rediscuss
  the ground and lowest excited potential energy curves (PECs) of H-He
  and the electric transition dipole moments (Sect.~\ref{sec:potHHenoSO}),
 with a stronger focus on  their long-distance behavior.
 A detailed correlation to the dissociated atomic states and its effect on the 
 transition dipole moments is specifically discussed in
 Sect.~\ref{sec:hhepec}. We consider spin-orbit (SO) coupling  in Sect.~\ref{sec:potHHeSO}.

A similar  asymmetry as in the Lyman-$\alpha$ H-He line profile
   also exists in
 the resonance broadening of hydrogen perturbed by collisions with H atoms.
In our first calculations for the resonance broadening of hydrogen
perturbed by collisions with H and H$^+$ \citep{allard1994}, we were mainly
interested in quasi-molecular absorption of  transient H$_2$ and
H$_2^+$ molecules   in the  far red wing of Lyman-$\alpha$.
Singlet states of H$_2$ lead to  line satellites
 from the free-free transitions
\hbox {$B^1\Sigma_u^+$ $\rightarrow$  $X^1\Sigma_g^+$} and 
\hbox {$C^1\Pi_u$ $\rightarrow$  $X^1\Sigma_g^+$}.
 These states also are responsible for the bound-bound 
 Lyman and Werner H$_2$ bands.
Triplet states only lead to a blue asymmetry because of
a close line satellite that appears as a shoulder in the blue wing.
These improved theoretical calculations of the complete Lyman-$\alpha$ profile
including both red and blue wings were applied
to the interpretation of {\it International Ultraviolet Explorer}  ({\it IUE}) 
and {\it HST} spectra.
They were shown to be 
fundamental in the interpretation of UV  spectra of variable
DA white dwarfs (ZZ Ceti stars) \citep{koester1994}. 
The analysis of the Lyman-$\alpha$ satellites in the far red wing
is not only a way to establish
the location of the ZZ Ceti instability of variable
DA white dwarfs,  but also a test of the assumptions
about convection efficiency \citep{bergeron1995}.

The  correlation diagram for H$_2$ states
contributing to Lyman-$\alpha$ shown in Table~2 of  \citet{allard1994}
was not correct because of an error in the preliminary ab initio
calculations of the 
$^3\Sigma_g^+$-$^3\Sigma_u$ transition moments.
This error was noted in  \citet{allard1998b}, who
pointed out that the variation of the radiative dipole moment must be
included in the line profile calculation.
  A new correlation diagram was presented in  \citet{allard2009a}
to correct Table~2 of \citet{allard1994}. This correlation diagram
has been used in \citet{allard1998b} and in our subsequent work. 
Electronic transition moments among singlets and triplets 
computed   by \citet{spielfiedel2003} and \citet{spielfiedel2004} were used
in  \citet{allard2009a} for
 an exhaustive study of the red wing of Lyman-$\alpha$ in order 
to determine the contribution of the triplet transition 
$b^3\Sigma_u^+ \rightarrow a^3\Sigma_g^+$ that was not considered in \citet{allard1994}.
 Although we never clarified the correct contribution of
 triplet states  to the blue wing  of the Lyman-$\alpha$ line, 
a blue line satellite was observed in
experimental spectra  \citep{kielkopf1995,kielkopf1998}.

We  discuss the effect of the $2s-2p$ degeneracy lifting on the long-distance behavior of H$_2$ states in Sect.~\ref{sec:potHH}. In Sect.~\ref{sec:lyHHHe}  we present a  study of the blue wing of the Lyman-$\alpha$ line
perturbed by collisions with H and He atoms in order to examine their relative contributions  in 
  the Lyman-$\alpha$ spectrum.

\section {Diatomic potentials and electronic transition dipole moments}

The transient interactions of atoms during radiative collisions are the main physical
quantities needed for a good understanding of the effect of collisional processes on radiative transfer in stellar
atmospheres and the spectra emitted by white dwarf stars.
We consider herafter H-He with and without spin-orbit coupling and H-H.

\subsection{H-He without SO coupling}
\label{sec:potHHenoSO}

The ab initio calculations of the potentials  
 were carried out with the MOLPRO package \citep{molpro2015} using  a very large Gaussian basis set that was initially taken from the $spdfgh$ aug-cc-pV6Z  basis 
set  of the MOLPRO library  for both He and H atoms, complemented by additional diffuse functions. 
On  both He and H,  the aug-cc-pV6Z  Gaussian basis set was complemented  by   diffuse functions in each $l$ manifold.
The added exponents  on He were the same as in the  previous work of \citet{allard2020}.
A slightly larger complementary set of  functions  was determined and used on H in order to accurately describe the atomic spectrum of hydrogen up to $n$=4, with
both diffuse functions and sometimes intermediate exponents improving  the nodal structure (Table~\ref{tab:basis} of the Appendix).
Thus the  basis set includes 239  Gaussian functions on He and 297 on H.

\begin{table}
\centering
\vspace{8mm}
\begin{tabular}{c|c|c|c}
\hline\\
        $level$  & Coulomb /DKH&Coulomb/DKH/mass & experimental\\
\hline\\
        1s&0 &0&0\\
        2s &82303.923&82259.124&82258.954\\
        2p$_{3/2}$&& & 82259.285 \\
        2p$_{1/2}$ &&& 82258.917 \\
        2p &82304.240&82259.440 & 82259.163\\
        3s &97545.453& 97492.357& 97 492.222  \\
        3p &97545.607 & 97492.511& 97492.293\\
        3d &97545.773 & 97492.678& 97492.341\\
        4s &102880.238 &102824.234 & 102823.853\\
        4p &102880.304 &102824.304 &102823.882\\
        4d &102880.366 &102824.368 &102823.903\\
        4f &102880.882 &102824.881 &102823.914\\
\hline\\
\end{tabular}
        \caption{ Atomic energy levels of hydrogen (in cm$^{-1})$: theoretical levels including the DKH contribution (second column), theoretical levels with DKH contribution and finite proton mass correction (third column), experimental values (fourth column) taken from  \citet{nist2020}. Experimental values  generally indicate  the weighted average over the spin-orbit components, except for the 2$p$ level, for which fine structure is explicitely detailed.}
\label{tab:hlevel}
\end{table}

A unique set of molecular orbitals was obtained from a relativistic Hartree-Fock (RHF) calculation of HeH$^+$, incorporating the scalar
relativistic effects (Darwin and mass-velocity contributions) within the Douglas-Kroll-Hess (DKH)   scheme \citep{reiher2006,nakajima2011} at second
order. The virtual orbitals of HeH$^+$ provide 
relevant excited molecular orbitals that properly dissociate into the exact
orbitals  of H
(as represented  within the present basis).  All subsequent calculations
  include the DKH contributions. The configuration interaction (CI)  was
  generated by a primary complete active space (CAS) 
including 14, 7, 7, and 4 molecular orbitals
in each of the a$_1$, b$_1$, b$_2$, and $a_2$ manifolds of the C$_{2v}$ point group. This ensured that all $\Sigma^+$, $\Pi$, $\Delta,$ and $\Phi$ states dissociating up to n=4 were properly
described in the CAS space. Finally,  an 
MRCI calculation \citep{knowles92}
generated from this CAS space was conducted for the $A_1$, $B_1$ , and $A_2$ manifold
with 13, 7, and 3 eigenstates, respectively. Although we  only discuss 
the Lyman-$\alpha$ contribution here, the relevant electric dipole transition moments corresponding to Lyman and Balmer transitions
up to $n$=4 were determined  in the same calculations and will be the object of future publication.  Relativistic effects are obviously very tiny on hydrogen. Nevertheless, because all excited levels present degeneracy in the simple nonrelativistic Coulomb scheme, it may be important for a correct long-range description to account even for very tiny splitting.

The DKH contribution, which is essentially active in the inner parts of the wave functions close to the atoms,   splits the degenerate components of a given $n$  and stabilizes more significantly the low  $l$ levels. It also slightly increases the excitation  energies from the 1s level. In the present work that is concerned
with the 2$s$, 2$p$ states,  these scalar contributions were complemented
by  accounting for spin-orbit coupling with similar magnitude.
For all levels of H $n$=1-4, the use of the extensive basis set and inclusion of the DKH correction provides 
 transition energies  obtained with an  accuracy better than 0.5 cm$^{-1}$ compared to the experimental atomic data ( \citet{nist2020}). Table~\ref{tab:hlevel} illustrates  the achieved accuracy (in particular, the  ordering of the $2s$,$2p$ states)  that is important to determine the configurational mixings that span the
molecular states and the  resulting transition probabilities that contribute to  Lyman-$\alpha$.

\begin{table}
        \centering
        \begin{tabular}{cc}
        \hline
              & \\
        Dissociation &  molecular state  \\  
              & \\
        \hline
        He(1s$^2$)+H(1s) &    $X^2\Sigma^+$\\
        He(1s$^2$)+H(2s,2p) &    $C^2\Sigma^+$, $A^2\Sigma^+$, $B^2\Pi$\\
              & \\
        H(1s)+H(1s) &    $X^1\Sigma_g^+$, $b^3\Sigma_u^+$\\
        H(1s)+H(2s,2p) &    $EF^1\Sigma_g^+$, $B'^1\Sigma_u^+$, $h^3\Sigma_g^+$, $e^3\Sigma_u^+$\\
                  &    $GK^1\Sigma_g^+$, $B^1\Sigma_u^+$, $a^3\Sigma_g^+$,\\ 
                    & $f^3\Sigma_u^+$, $I^1\Pi_g$, $C^1\Pi_u$, $i^3\Pi_g$, $c^3\Pi_u$\\
              & \\
        \hline
        \end{tabular}
        \caption{Symmetry  and labeling of  molecular states dissociating into H(2s,2p)+He and H(2s,2p)+H }
        \label{tab:cor}
\end{table}

\begin{figure}
\centering
\vspace{8mm}
\includegraphics[width=8cm]{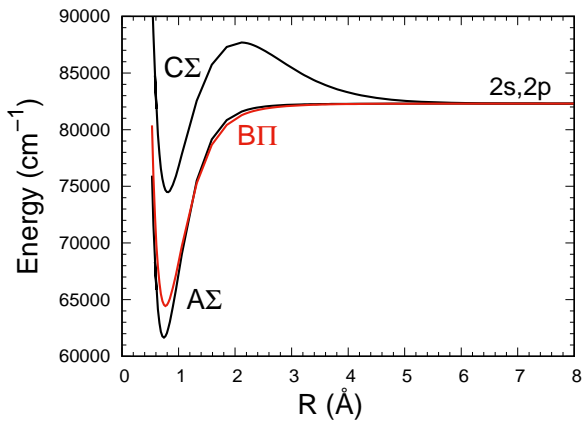}
        \caption{MRCI adiabatic potential energy curves of H-He correlated with the 2$s$, 2$p$ atomic levels.
        The zero energy corresponds to H(1s)+He(1s$^2$) at dissociation.}
\label{fig:e2}
\end{figure}

  The symmetry and labeling of the molecular states that dissociate into He(1s$^2$) + H(2s,2p)  is shown in Table \ref{tab:cor}, and their potential energy curves are plotted in  Fig. \ref{fig:e2}. States $A\Sigma$ and $B\Pi$ are attractive with
a minimum located around R=~0.75 \AA\/. State  $C^2\Sigma^+$ 
presents a repulsive character at medium and long range, resulting in a barrier
and a shallower short-distance well depth.
This medium-range repulsion  is associated with the repulsion between the electronic density  of the $2s$ and $2p_\sigma$  hydrogen orbitals along the axis and the electrons of the
helium atom.

\subsection{H-He PECs and correlation to dissociated atomic states}
\label{sec:hhepec}

\begin{figure}
\centering
\vspace{8mm}
\includegraphics[width=8cm]{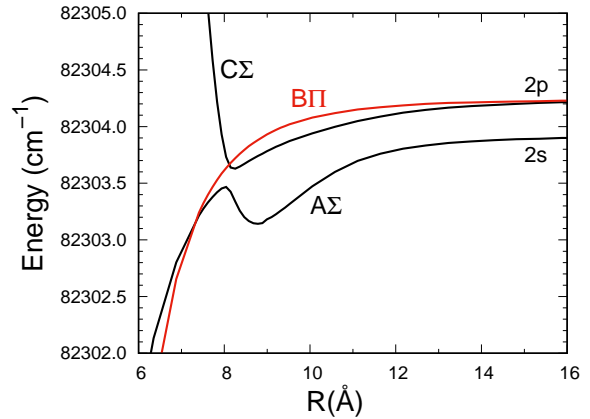}
        \caption{ Long range zoom of the MRCI adiabatic H-He potential energy curves of  states  $A$, $B$ and $C$ dissociating into  H(n=2)+ He.  The zero energy corresponds to H(1s)+He(1s$^2$) at dissociation.}
\label{fig:elr}
\end{figure}

 For distances smaller than 8.1 $\AA$, the results we show
in Fig.~\ref{fig:elr} ~do not differ substantially
from the recent calculation of \citet{allard2020}, which was essentially achieved   with  a similar  methodology.
We used  the same   aug-cc-pV6Z basis set complemented  with diffuse functions (the same functions for helium, and slightly more diffuse functions on hydrogen), and a larger  CAS space generating the MRCI (three electrons in 
32 orbitals, instead of three electrons in 14  orbitals in the former work). None of these differences are expected to provide significant quantitative changes  at  short distance concerning the ground state and the states dissociating into H(n=2)+He.  However,  no explicit
mention of the relativistic effects was   made \citep{allard2020},
and the $C$ state potential energy curve was found to converge to
the unique Coulomb 2s/2p asymptote  with a  dipole moment for the $X-C$ transition that vanished asymptotically, meaning an adiabatic correlation of the obtained $C$ state with  the dipole-forbidden $2s$ atomic state.  Conversely, 
the transition dipole moment  $X-A$  was found to converge to that of the atomic $1s-2p$ transition dipole moment value. 

We here focus  strongly on the medium and long distance of the potential curves. In particular,
at the asymptotic limit, the DKH relativistic correction lifts the
hydrogen 2$s$/2$p$ degeneracy, 
lowering the 2$s$ state more than the 2$p$ state. This results in a 2$s$-2$p$ splitting of 0.316 cm$^{-1}$. As a result, a 
long-distance avoided crossing around 8.1 $\AA$  occurs in the $^2\Sigma^+$
manifold
between the lower adiabatic state correlated with 2$s$  and the upper state
correlated with 2$p$. The state correlated with $2s$ has a tiny well at 8.7 $\AA$.  This causes the adiabatic upper state (labeled $C$ at short distance by 
spectroscopists) to correlate with the $2p$ asymptote, while state $A$ is
correlated with 2$s$. 
The examination of the dipole transition moments  from the ground state,  shown in Fig.~\ref{fig:dip2}, 
reveals a crucial implication.
This long-range avoided crossing induces a  kink at 8.1 $\AA$
in the dipole moment $X-C$ and a sign change in   the $X-A$ dipole transition
moment.
Moreover,  both  adiabatic  states have   transition moments that stabilize around  0.4-0.6 au below R=8.1~\AA\/, which means that
the upper state loses some $2p$ character while the lower gains it,
regardless of the avoided crossing at 8.1~\AA\/.
Thus, the situation results from 
a gradual  $2s/2p$ mixing  increasing with short distance, superimposed with  the  sharp  avoidance  at 8.1~\AA\/.

\subsection{H-He with SO coupling}
\label{sec:potHHeSO}

 SO coupling was incorporated 
following  the atom-in-molecule-like scheme introduced by 
\citet{cohen1974}. 
 It  relies
on an effective monoelectronic spin-orbit coupling operator,
\begin{equation}
  H_{SO}=\sum_i h_{SO}(i)=\sum_i\zeta_i \hat{l}_i . \hat{s}_i \; ,\end{equation}
 where $\hat{l}$ and $\hat{s}$ are the orbital and spin-moment operators, and $\zeta_i$ is the effective spin constant associated with a given atomic electronic shell.
  
\begin{figure}
\centering
\vspace{8mm}
\includegraphics[width=8cm]{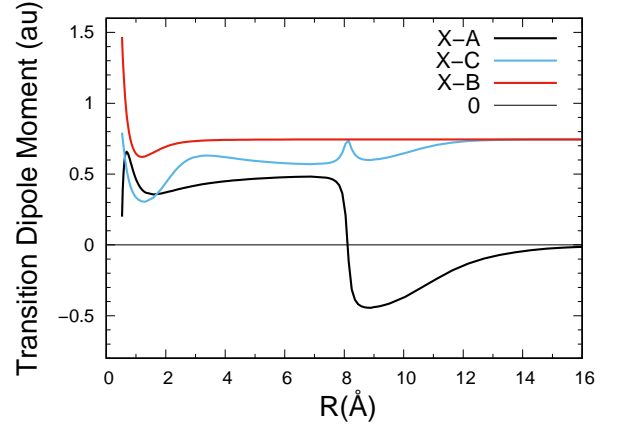}
\caption{Spin-orbit-less transition dipole moments of H-He from the $X$ ground state toward the adiabatic $A$, $B,$ and $C$ states of 
         H-He.} 
\label{fig:dip2}
\end{figure}

 The total Hamiltonian $H_{el}+H_{SO}$ is expressed in the basis set of the
 eigenstates (here with total spin projection $\sigma=\pm\frac{1}{2}$) of the purely electrostatic
 Hamiltonian $H_{el}$.
 Because the core electrons of He define a closed shell, the
 spin-orbit coupling between the molecular  many-electron doublet
 states $\Psi_{k\sigma}$, approximated at this step as
single  determinants with the same  closed shell $1s\sigma^2$ as the He subpart, 
is isomorphic to that
between  the  singly occupied molecular spin-orbitals $\phi_{k\sigma}$,
which are asymptotically correlated with the six  $2p$  spin-orbitals and two $2s$
spin-orbitals of H.
In its original formulation, the Cohen and Schneider approximation  consists
of assigning these   matrix elements to their asymptotic (atomic) values. However, in the case of HeH,
the  $2s$ and $2p$ shells are asymptotically degenerate, which offers a favorable situation
for electronic mixing. As previously mentioned, both adiabatic states $\Psi_{2s\Sigma}$ and $\Psi_{2p\Sigma}$  depart  from their asymptotic character.
Conversely, the $\Psi_{2p\Pi}$ state   does not significantly 
mix with  any other states and essentially conserves the asymptotic transition dipole moment 
from the ground state shown in Fig.~\ref{fig:dip2}.

In case of strong mixing, application of the  Cohen and Schneider
scheme is more relevant in a basis of diabatic or quasi-diabatic states,
as has been discussed in previous works \citep{allard2020}.
Thus we determined $\Phi_{2s\Sigma}$ and $\Phi_{2p\Sigma}$ diabatic states
through a unitary transform  of the adiabatic states $\Psi_{2s\Sigma}$ and
$\Psi_{2p\Sigma}$
and the constraint that the transition dipole moments from the ground state $X~1^2\Sigma^+$ to the quasi-diabatic states at finite distance remain as close as possible
to their asymptotic values, namely zero and 0.744 (asymptotic limit of the MRCI calculation),

\begin{equation}
        \begin{pmatrix}
                \Phi_{2s\Sigma}\\
                \Phi_{2p\Sigma}\\
        \end{pmatrix}
        =
        \begin{pmatrix}
        \cos \theta& -\sin \theta\\
        \sin \theta& \cos \theta
        \end{pmatrix}
        \begin{pmatrix}
                \Psi_{2s\Sigma}\\
                \Psi_{2p\Sigma}\\
        \end{pmatrix}
.\end{equation}

Fig.~\ref{fig:dip2} shows that  the transition
dipole  moments to the quasi-diabatic
 states are  now exactly zero for state $\Phi_{2s\Sigma}$ and
almost constant for state $\Phi_{2p\Sigma}$.
The two latter  and state $\Phi_{2p\pi}=\Psi_{2p\Pi}$ considered as
diabatic therefore
span the Cohen and Schneider spin-orbit matrix, the electronic Hamiltonian  being nondiagonal in
the $\Sigma$ states manifold, 
\begin{equation}
  <\Phi_{k\sigma}|H_{SO}|\Phi_{l\tau})>= 
  <\phi_{k\sigma}({\infty})|h_{SO}|\phi_{l\tau}({\infty})> 
.\end{equation}

 The effective spin-orbit  constant  of the 
$2p$ shell of hydrogen is taken empirically as 2/3 times the spin-orbit 
splitting 0.366 cm$^{-1}$, namely $\zeta_{2p}$= 0.244 cm$^{-1}$ \citep{nist2020}. 
The total Hamiltonian matrix  $H=H_{el}+H_{SO}$  for the n=2 manifold is
thus an  8$\times$ 8 complex matrix that can  decouple into the real
matrices  according to different values  of $\Omega=M_l+M_s$,
where $M_l$ and $M_s$ are the  orbital and spin moment projections on the molecular axis. For $\Omega=\pm 1/2$ and $\Omega=\pm 3/2$, these matrices read as follows:

\begin{displaymath}
        \Omega=\pm1/2\hspace*{1cm}
\left(
\begin{array}{ccc}
E_{2s\Sigma}&V&0\\
        V&E_{2p\Sigma}&\zeta\sqrt{2}/2\\
        0&\zeta\sqrt{2}/2 &E_{2p\Pi}-\zeta/2\\
\end{array}
\right)
\end{displaymath}
\begin{displaymath}
        \Omega=\pm 3/2\hspace*{1cm}
\left(
\begin{array}{c}
        E_{2p\Pi}+\zeta/2\\
\end{array}
\right).
\end{displaymath}

Here $E_{2s\Sigma}$ and $E_{2p\Sigma}$ are the energies of the quasi-diabatic states and $V$ is their electronic coupling. $E_{2p\Pi}$ is the energy of the  $\Pi$ state.
 The diagonalization of the above matrix at each
internuclear distance provides the spin-orbit  eigenstates and energies.
Despite being performed in an  extensive basis, our calculation cannot reach  sub
cm$^{-1}$ accuracy, which is required to investigate the molecular fine structure close
to the asymptote because the SO splitting is only 0.366 cm$^{-1}$. In particular,
the difference between the experimental atomic $2s$ state and the average of the  $2p$ states is 0.209 cm$^{-1}$, while the
calculation at separation yields 0.316 cm$^{-1}$.  Thus we shifted the 
nonrelativistic diabatic potential $E_{2s\Sigma}$  upward by 0.107 cm$^{-1}$.
As a consequence of SO splitting,  the  $2p_{1/2}$ level now lies closely below the $2s_{1/2}$ state,
by  0.034 cm$^{-1}$ in the calculation $\text{versus}$ 0.035 cm$^{-1}$ experimentally.

\begin{figure}
\centering
\vspace{8mm}
\includegraphics[width=8cm]{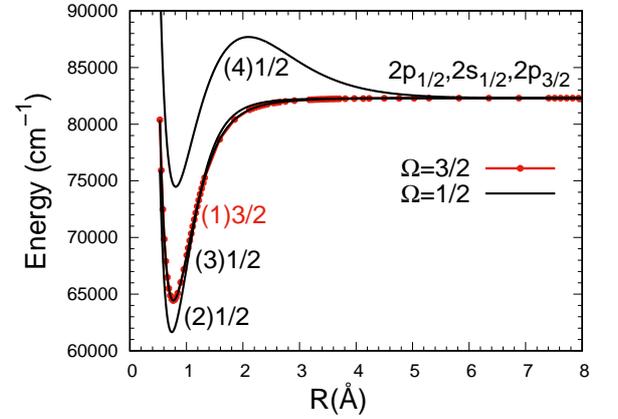}
        \caption{Adiabatic H(n=2)+He potential energy curves of  molecular states including SO coupling. For better
	display, the red line of state $(1)3/2$ state is shown with superimposed dots. 
        Zero energy corresponds to H(1s)+He(1s$^2$) at dissociation.}
\label{fig:eso}
\end{figure}

\begin{figure}
\centering
\vspace{8mm}
\includegraphics[width=8cm]{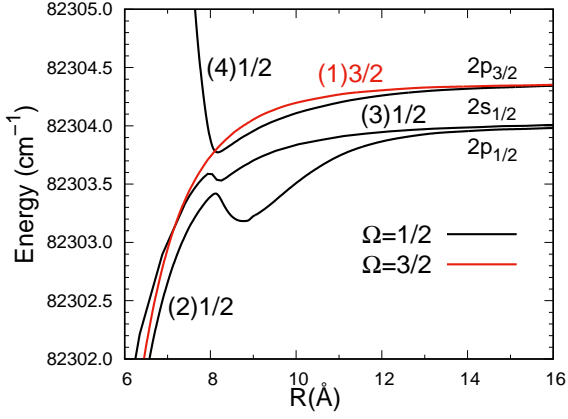}
        \caption{ Long-range zoom of the adiabatic potential energy curves of H(n=2)+He states including SO coupling.
        Zero energy corresponds to H(1s)+He(1s$^2$) at dissociation.}
\label{fig:esolr}
\end{figure}

\begin{figure}
\centering
\vspace{8mm}
\includegraphics[width=8cm]{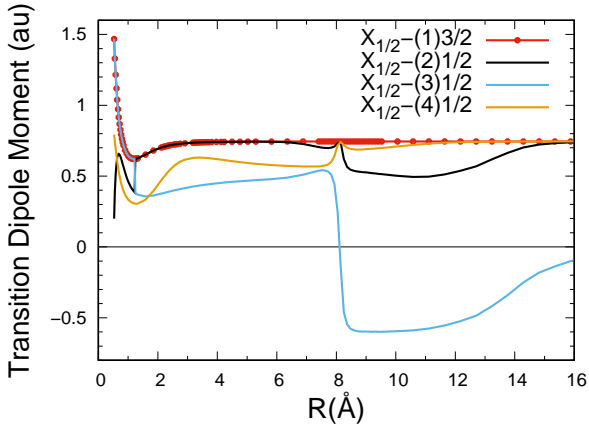}
	\caption{Transition dipole moments from the $X_{1/2}$ ground state toward the adiabatic states dissociating into  H(n=2)+He including SO coupling. For better display, the red line of the $X_{1/2}-(1)3/2$ transition dipole moment is shown with
  superimposed dots. } 
\label{fig:dipso}
\end{figure}

 The molecular eigentates can be  labeled in adiabatic order  in each  $\Omega$ manifold, 
 namely (2)1/2, (3)1/2, (4)1/2, and (1)3/2 above the ground state (1)1/2 (also named herafter  $X_{1/2}$). 
 They adiabatically  correlate   with atomic asymptotes  $2p_{1/2}$, $2s_{1/2}$, $2p_{3/2}$ , and $2p_{3/2}$, respectively. 
At short distance R$<1.2\AA$ , the lowest excited eigenstate (2)1/2 is  essentially spanned by  $A\Sigma$.
However at intermediate range 1.2-7.3~\AA, (2)1/2 switches to  $B\Pi$ which lies below $A\Sigma$. At R=$~8.1 \AA$  it  undergoes another avoided crossing and is finally adiabatically correlated with the lowest asymptote $2p_{1/2}$. 
Consistently, state (3)1/2 has a $B\Pi$ character at short distance,  switches to  $A\Sigma$ in the intermediate range
and is adiabatically correlated with asymptote $2s_{1/2}$. Finally state (4)1/2 has essentially 
a $C\Sigma$ character for R$< 8.1 \AA$, however as a consequence of the avoided crossing between the $\Sigma$ state at 8.1~$\AA$ , it becomes  adiabatically correlated with asymptote $2p_{3/2}$.  Only state $(1)3/2$  remains identical to its parent state $B\Pi$ at all distances, except for an asymptotic shift. It should be noted that in a diabatic picture where the states  may  cross (but where the  hamiltonian is no longer diagonal), three 1/2 states keeping the $A\Sigma$, $B\Pi$ and $C\Sigma$ character almost up to dissociation could be defined, correlated to $2p_{1/2}$, $2p_{3/2}$ and $2s_{1/2}$ respectively after multiple crossings.

The  transition  dipole moments (Fig.~\ref{fig:dipso}) between the spin-orbit  
states $\Psi^{SO}_m$
can easily be determined from those computed between the 
quasi-diabatic CI wavefunctions,

 \begin{equation}
\mathbf{D}^{SO}_{mn}=< \Psi^{SO}_m|\mathbf{D}|\Psi^{SO}_n>=\sum_{k\sigma,l\tau}c^m_{k\sigma}
c^n_{l\tau}<\tilde{\Phi}_{k\sigma}|\mathbf{D}|\tilde{\Phi}_{l\tau}>\delta_{\sigma\tau}.
 \end{equation}

 The  dipole moments of transitions $X_{1/2}-(2)1/2$ and $X_{1/2}-(4)1/2$ show kinks at 8.1 $\AA,$ while that of the $X_{1/2}-(3)1/2$ transition changes sign. This feature results  from  the avoided or actual crossings between their three $A$, $B,$ and $C$  parents
 and the consecutive
multiple crossings of the spin-orbit states at this distance. The transition moment $X_{1/2}-(1)3/2$ remains clearly equal to that of  $X-B$ . The dipole moments of transitions $X_{1/2}-(2)1/2$ and $X_{1/2}-(3)1/2$ furthermore show a sudden exchange at 1.2 $\AA$ due to the crossing of their  parent states $A\Sigma$ and $B\Pi$, respectively, at that distance.

 The  electronic structure of HeH involving fine structure has also been investigated with
pseudopotentials \citep{kielkopf2021}, in excellent agreement with our calculation in the region R>6 $\AA$.  Although we carried out an analysis of the effect of spin-orbit
coupling on the H-He potential curves and transition moments, spin-orbit
coupling is not taken into account in the collisional section below.
The most important  effect, namely  the adiabatic correlation
of the $C$ state with the upper allowed $2p$ asymptote associated with a nonvanishing $X-C$ dipole
moment, is  maintained  when spin-orbit
coupling is accounted for:  the $C_{1/2}$ state, spanned by  the parent state $C\Sigma$  at short distance,  correlates with the dipole-allowed atomic  state  $2p_{3/2}$, and the   $X_{1/2}-(4)1/2$ transition dipole remains  finite at large distance.

\subsection{H-H potentials}
\label{sec:potHH}

\begin{figure}
\centering
\vspace{8mm}
\includegraphics[width=8cm]{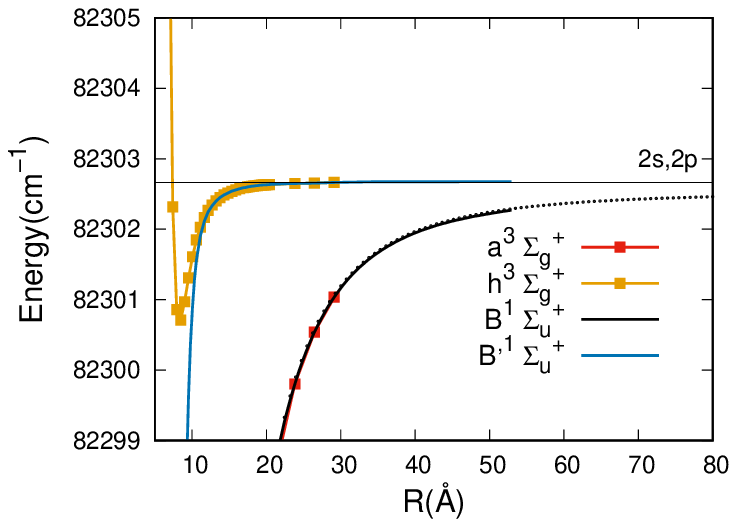}\\
\includegraphics[width=8cm]{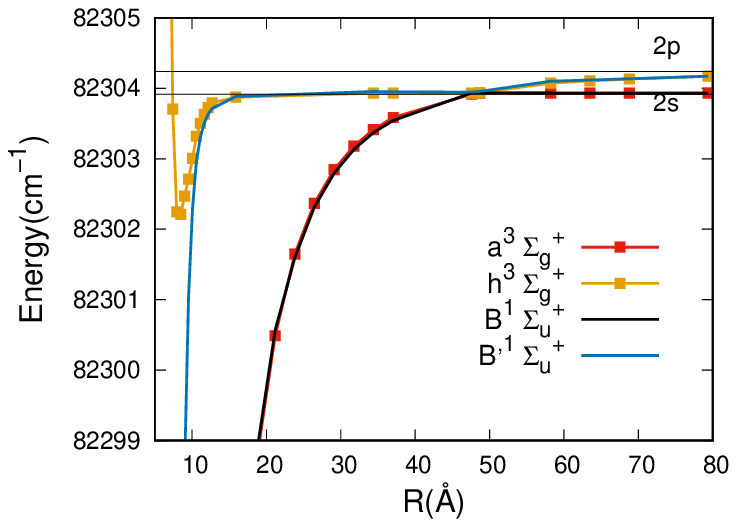}
        \caption{ Long-range zoom of the adiabatic potential energy curves of  states  $a^3\Sigma_g^+$, $h^3\Sigma_g^+$, 
        $B^1\Sigma_u^+$ , and $B'^1\Sigma_u^+$ of H+H(n=2). Top: Data of Spielfiedel et al. (2001, private communication) without scalar relativistic terms.
        Bottom: Our  CI results including the DKH correction.
        Zero energy corresponds to H(1s)+H(1s) at dissociation. The dotted line corresponds to the $C_3/R^3$ asymptotic extrapolation
        (hidden within the calculated curves lines  in the bottom plot).}
\label{fig:hhlr}
\end{figure}

 In \citet{allard1994} the theoretical potentials for the binary
interaction of one hydrogen atom with
another hydrogen atom were taken from the calculations of
\citet{sharp1971} and \citet{wolniewicz1988}. The  dependence of the
probabilities  of the  allowed molecular transitions on internuclear
separation  contributing
to the Lyman-$\alpha$  line were taken from 
\citet{dressler1985} for the singlet states and preliminary
ab initio results  for the transitions between the triplet states.
The allowed transitions contributing to Lyman-$\alpha$ were summarized in Table~4 of \citet{allard1994}, but the  labels of the $a$ and
$h$ triplet states were interchanged. 
  The symmetry and labeling of the H-H states  are
listed in Table~\ref{tab:cor}.
It might be wondered whether  the inclusion
of relativistic terms might seriously affect  the long-distance behavior
of the potentials and the transition dipole moments, as  found in HHe.
In Fig.~\ref{fig:hhlr} we show  the long-range behavior of the theoretical
quasi-full CI H-H potentials  of Spielfiedel (2001, private communication) 
carried out in the Coulomb framework, and MRCI calculations (full CI)
conducted  on H-H with the same basis set as was used
on hydrogen in HHe (see above) and the DKH correction. 
The splitting of the $2s$ and $2p$ levels also induces
avoided crossings in H$_2$. While states $h^3\Sigma_g^+$ and
$B'^1\Sigma_u^+$ join the asymptote around 15~\AA\,  states $a^3\Sigma_g^+$
and $B^1\Sigma_u^+$ remain  significantly attractive up to very large
separation because of the $1/R^3$ contribution. They exhibit  avoided
crossings
(within each symmetry manifold) with the former states (diabatically
correlated with atomic forbidden asymptotes).  However, in H$_2$, these
avoided  crossings take place at very long distance,
namely around $R$=45~\AA\/. At this interatomic separation, the electronic
coupling vanishes and we can consider that in a collisional
situation the system will follow a diabatic behavior, namely it will move along
diabatic potentials that cross. In the present  case with vanishing
coupling,  diabatic potentials   can easily be defined  by reassigning the adiabatic potential values  before and after the (avoided) crossing.

  Because only the difference potentials are meaningful for the treatment of
  collisional broadening, we can infer that  scalar relativistic corrections are not
  needed and the potentials in the Coulomb description can be used simply as
  diabatic potentials.  We therefore
  used  the theoretical H-H potentials
 of Spielfiedel (2001, private communications) together with the 
electronic transition moments among singlets and triplets taken from 
\citet{spielfiedel2003} and \citet{spielfiedel2004}.
As with H-He, spin-orbit coupling is not
considered in the collisional section that follows.

\section{Collisional profiles perturbed by neutral H and He atoms}
 \label{sec:lyHHHe}

  Quasi-molecular lines  in the red wing of
  Lyman-$\alpha$, Lyman-$\beta$, and Lyman-$\gamma$ 
arise from radiative collisions of excited atomic hydrogen with
unexcited neutral hydrogen atoms or protons
\citet{allard1991,allard1998a,allard1998b,allard2000,allard2004a,allard2004b,allard2009b}.
A general unified theory in which the electric dipole moment varies
during a collision \citep{allard1999} is  essential 
as the blue H-H quasi-molecular line satellite  is 
due to an asymptotically forbidden transition
\mbox{ $b \, ^3\Sigma_{u} \rightarrow h\,^3\Sigma_{g}$}.   

\subsection{Unified theory}

Starting with the \citet{anderson1952} theory
suitably generalized to include degeneracy,
a unified theory of spectral line broadening 
\citep{allard1999}
was developed to calculate neutral atom spectra, given the
interaction and the radiative transition moments of relevant states
of the radiating atom with other atoms in its environment.
The unified profiles  are the Fourier 
transforms (FT) of the autocorrelation functions as given by Eq.~(121) 
of~\citet{allard1999}, in which the contributions from the different 
components of a transition enter with their statistical weights.
A pairwise additive assumption allows us to calculate 
$I(\Delta \omega)$, when
$N$ perturbers interact as the FT of the $N^{\mathrm{th}}$ power of the
autocorrelation function $\phi(s)$ of a unique atom-perturber
pair.
For a perturber density $n_p$, we obtain
\begin{equation}
\Phi(s) = e^{-n_{p}g(s)} \; ,
\label{eq:intg}
\end{equation}
where the decay of the autocorrelation function with time leads to atomic line
broadening.  

For a transition \mbox { $\alpha =(i,f)$} from an initial state~$i$ 
to a final state~$f$, we have 
\begin{eqnarray}
g_{\alpha}(s) && \,= \frac{1}
{\sum_{e,e'} \, \! ^{(\alpha)} \, |d_{ee'}|^2 }
\sum_{e,e'} \, \! ^{(\alpha)} \; \; \nonumber \\ 
&&  \int^{+\infty}_{0}\!\!2\pi\rho d\rho
\int^{+\infty}_{-\infty}\!\! dx \; 
\tilde{d}_{ee'}[ \, R(0) \, ] \, \nonumber \\ 
&&[ \, e^{\frac{i}{\hbar}\int^s_0 \, dt \;  
V_{e'e }[ \, R(t) \, ] } \,
\, \tilde{d^{*}}_{ee'}[ \, R(s) \, ] \, - \, \tilde{d}_{ee'}
[ \, R(0) \, ] \, ]
\; . \;
\label{eq:gcl}
\end{eqnarray}

 In Eq.~(\ref{eq:gcl}) the symbols $e$ and  $e'$  label
the energy surfaces  on which the interacting
atoms  approach the initial and final atomic states of the transition. 
The sum $\sum_{e,e'} ^{(\alpha)}$ is over all pairs ($e,e'$)  such that
\mbox{$\omega_{e',e}(R) \rightarrow \omega_{\alpha}$} as 
\mbox{$R \rightarrow \infty$}.
 In the equations that follow, we review the meaning of these terms 
  in Eq.~(\ref{eq:gcl}).
In our context, the perturbation of the frequency of the atomic
transition during the collision results in a phase shift, 
$\eta(s)$,  which is calculated along a classical path $R(t) $  
that is assumed to be rectilinear.
At time $t$ from the point of closest approach, the atoms are separated by 
\begin{equation}
R(t) = \left[\rho ^2 + (x+\bar{v} t)^2 \right]^{1/2} \; , \;
\end{equation}
     where $\rho$ is the impact parameter of the perturber trajectory,
      $\bar{v}$ is the relative velocity, and 
$x$ the position of the perturber along its trajectory at time
$t=0$. 
We have for the phase term in  Eq.~(\ref{eq:gcl})
\begin{equation}
\eta(s)=  \frac{i}{\hbar}\int^s_0 \, dt \;
V_{e'e }[ \, R(t) \, ]
\label{eq:phase}
,\end{equation}
where $\Delta V(R)$, the difference potential, is given by
\begin{equation}
\Delta V(R) \equiv V_{e' e}[R(t)] = V_{e' }[R(t)] - V_{ e}[R(t)] \; ,
\label{eq:deltaV}
\end{equation}
and represents the difference between the  energies
of the quasi-molecular transition. The potential energy for a state $e$ is 
\begin{equation}
V_{e}[R(t)] = E_e[ R(t) ]-E_e^{\infty} \; .
\label{eq:V}
\end{equation}

\begin{figure}
 \centering
\resizebox{0.46\textwidth}{!}
{\includegraphics*{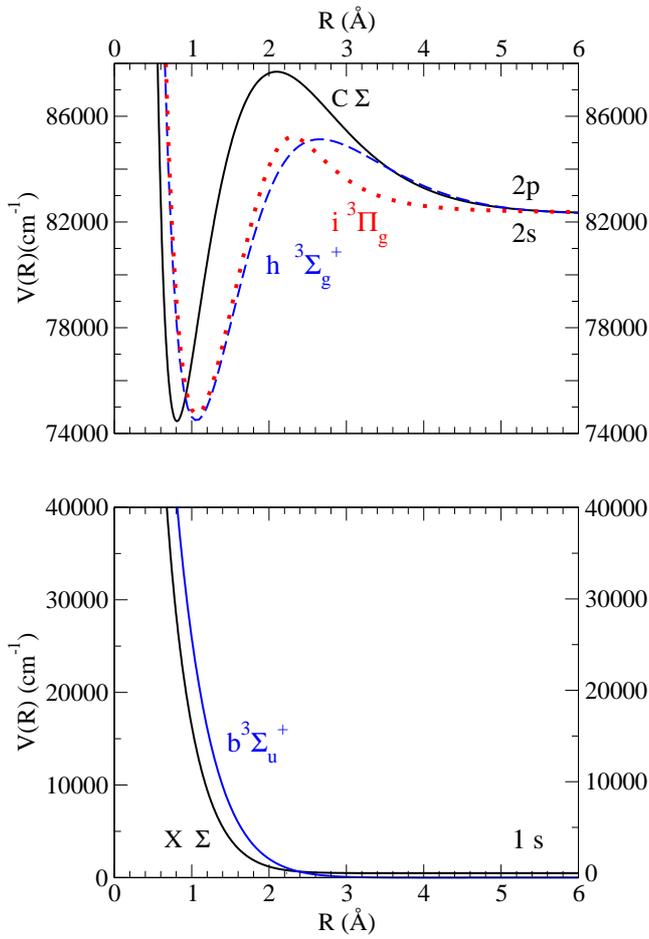}}
        \caption  {Top: Short-range part of the repulsive potential curve $C \Sigma$ (black line)
   of the H-He molecule compared to the 
   $h$ $^3\Sigma_g^{+}$ (dashed blue line) and $i$ $^3\Pi_g$ (dashed red line)
   states of the H$_2$ molecule.
           Bottom: $X$ (black line) and $b$ $^3\Sigma_u^{+}$ (blue  line).}
\label{sec:pots}
\end{figure}

 \begin{figure}
 \centering
\resizebox{0.46\textwidth}{!}
{\includegraphics*{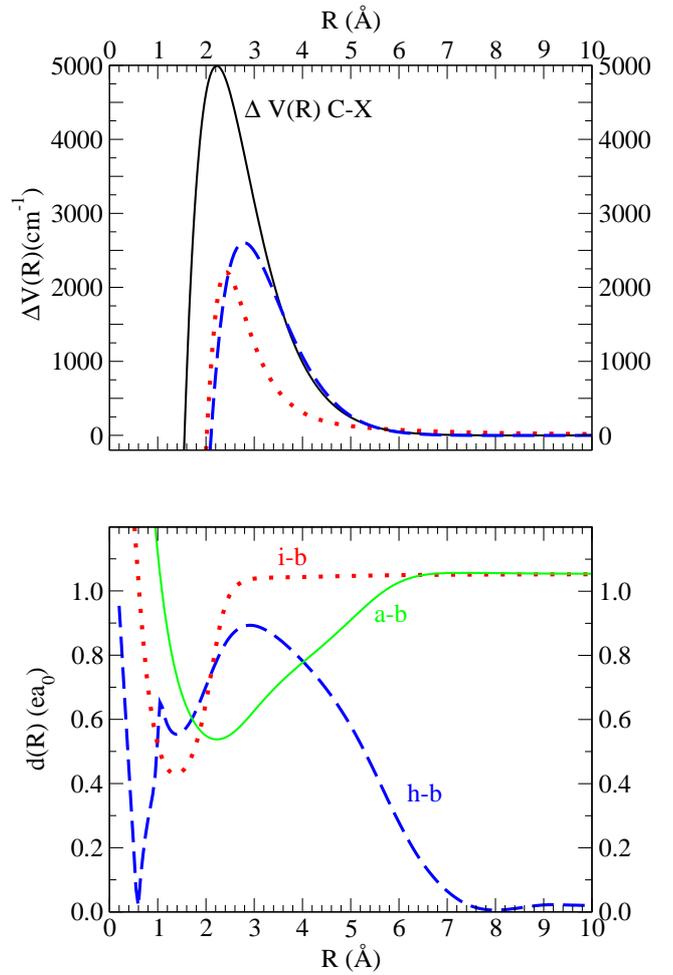}}
\caption  {Top: Difference  potential for the triplet transitions
  $h-b$ (dashed blue line), $i-b$  (dotted red line) of H-H
  compared to the $X-C$ transition of H-He.
  Bottom: Electric dipole transition  moments   between the
  triplet  states of H-H. The $h-b$  transition 
    is forbidden asymptotically as it is a transition between the $2s$
    and $1s$ states of the free hydrogen atom at large $R.$}  
\label{sec:vdiff}
 \end{figure}

 \subsection{Satellite bands due to H-H and H-He}
 
Within the assumption of additive superposition of interactions, we can
write the total profile as the convolution of the individual profiles
corresponding to perturbations by H and He,
\begin{equation}
 I_{tot}(\Delta \omega) =  I_{H-H}(\Delta \omega) * I_{H-He}(\Delta \omega).
\end{equation}

The  unified theory  \citep{allard1982} predicts that line satellites
 are centered
 periodically at frequencies corresponding to integer multiples of
 the extrema of  $\Delta V(R)$ (Eq.~(\ref{eq:deltaV})).
 However, their appearance depends on the
 value of the electronic dipole
 moments in the region of the maximum of $\Delta V(R)$  \citep{allard1998b}.

 The prediction of the shape  of the blue wing 
required us to  study the potential
energies of the two triplet transitions
\hbox {$b$ $^3\Sigma_u^{+}$  $\rightarrow$ $h$ $^3\Sigma_g^{+}$}
and \hbox {$b$ $^3\Sigma_u^{+}$  $\rightarrow$ $i$ $^3\Pi_g$}
that contribute  to the blue wing of the Lyman-$\alpha$ line
(Fig.~\ref{sec:pots}).
For comparison we overplot the potential energies of the  $X \, \Sigma$
and $C \, \Sigma$ states of the H-He system.
The prediction  of a line satellite in the blue wing of
the H-H and  H-He line profiles is related to the potential maximum at
short distance 
$R=2-3$~\AA\/ of the $C, \,h,$ and $i$ states.
This leads to a maximum
of the potential energy difference $\Delta V(R)$
for these transitions shown in Fig.~\ref{sec:vdiff}.
The electronic states $h$ and $b$ of the isolated radiator are not
connected by the dipole moment operator: $d_{hb}(R \rightarrow \infty) = 0$. 
As reported in \citet{spielfiedel2004},
the 2, 3 $^3\Sigma_g$ states (labeled $h$, $a$) undergo an avoided crossing at
equilibrium distance  and thus exchange their
character. The  radiative dipole   moment varies dramatically
with $R$ (bottom of Fig.~\ref{sec:vdiff}).
Allowed radiative transitions cannot occur between these two states,
but $d_{hb}(R)$ differs from zero when a perturber passes close
to the radiator.
Our  theoretical approach allows us to take
 this asymptotically forbidden transition 
\hbox { $b^3\Sigma_u^+ \rightarrow h^3\Sigma_g^+$} of quasi-molecular
hydrogen that dissociates into 
($1s, 2s$) atoms into account.
An other important factor  is the variation of
the dipole moment during the collision once modulated by the
Boltzmann factor $e^{-\beta V_e(r)}$ (Eq.~(117) of  \citet{allard1999}), 

\begin{figure}
 \centering
\resizebox{0.46\textwidth}{!}
{\includegraphics*{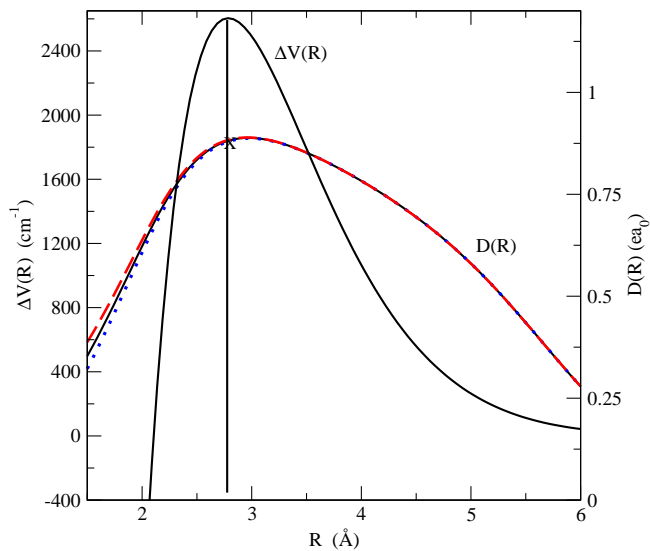}}
\caption  {Variation  with  temperature in  the   modulated dipole
  and the  difference potential of the triplet state $h$-$b$  of the
  H-H molecule. T=15000~K (dashed red line), T=12000~K (black line), and 
T=10000~K (dotted blue line).}
\label{sec:hb}
\end{figure}

\begin{equation}
  D(R)= \tilde{d}_{if}[R(t)] = d_{if}[R(t)]e^{-\frac {V_e [R(t)]}{2kT}} \; \; .
\label{eq:dip} 
\end{equation}
The Boltzmann factor $e^{-\frac {V_e (R)}{2kT}}$ in Eq.~(\ref{eq:dip})
appears because the
perturbing atoms  are in thermal equilibrium with the radiating
atom, which affects the probability of finding them initially at a given $R$.
In this case, where we consider absorption profiles due to triplet transitions,
$V_e$ is the $b$ $^3\Sigma_u^{+}$ ground-state potential.
In Fig.~\ref{sec:hb} we show $D(R)$ together with the corresponding
$\Delta V (R)$ for the $h$-$b$ transition.
The dipole moment $D(R)$ and the energy difference determining the transition wavelength
$\Delta V (R)$ are  maximum at
$R_{\rm ext}$= 2.8~\AA\/. In this instance, a
radiative transition is  induced by collisions.
Figure~\ref{sec:hb} shows that  $D(R)$ is  not    dependent
on temperature  throughout the
region where  the collision-induced  satellite is formed,
and therefore the blue wing of Lyman-$\alpha$
will not change with increasing temperature in the range of temperatures
10000-15000~K.

\begin{figure}
\centering
\vspace{8mm}
\includegraphics[width=8cm]{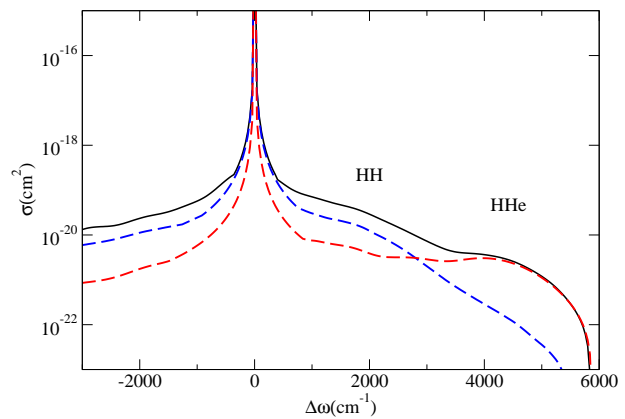}
\caption{Blue wings of Lyman-$\alpha$ due to H-H collisions
  (blue curve), H-He collisions (dashed red   curve), and simultaneous collisions
  by H and He (black curve).
  The H and He densities are 10$^{18}$~cm$^{-3}$, and the  temperature is 12000~K.}
\label{lymanHH18HHe18}
\end{figure}

 An examination of Fig.~\ref{sec:vdiff} leads us to expect
a farther blue satellite for H-He that arises from the extremum
of  5000~cm$^{-1}$ when the two atoms
are separated by about 2~\AA\/ compared to 2100-2500~cm$^{-1}$ for $i-b$ and
$h-b$.
Figure~\ref{lymanHH18HHe18} shows 
 the distinct wide satellite   at about 4200~cm$^{-1}$ owing
 to the $X$-$C$ transition, whereas the
 \hbox {$b$ $^3\Sigma_u^{+}$  $\rightarrow$ $h$ $^3\Sigma_g^{+}$} transition 
  yields a  blue shoulder centered  approximately at  1900~cm$^{-1}$ in 
  the blue wing  of  Lyman-$\alpha$ of  H-H.
  Although  $\Delta V (R)$ for the allowed transition 
  \mbox{$b \, ^3\Sigma_{u} \rightarrow i\,^3\Sigma_{u}$} has a
  maximum ({$\Delta V$= 2100~cm$^{-1}$}), it simply contributes to the
  blue asymmetry as it is blended in the near wing.

The wavelength of the theoretical collision-induced  satellite
is largely dependent on the
accuracy of the  difference potential of the two contributing states
 to the transition, while the strength of the absorption as a function 
of wavelength is dependent on  the radiative dipole 
moment  shown in Fig.~\ref{sec:hb}  and on the accuracy of the spectral
line shape theory.  
This line satellite has been observed experimentally in the
spectrum of a laser-produced plasma source, see Fig.~2 of
\citet{kielkopf1995} and Fig.~7 of \citet{kielkopf1998}.

\begin{figure}
\centering
\vspace{8mm}
\includegraphics[width=8cm]{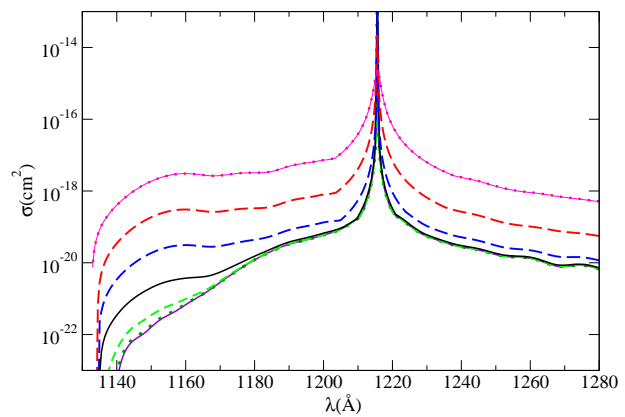}
\caption{Blue wing of Lyman-$\alpha$ simultaneously perturbed by
  He and H atoms for a different ratio of helium and hydrogen densities.
  From top  to  bottom,  $n_{\mathrm{He}}$/$n_{\mathrm{H}}$
  is $10^3$, $10^2$, 10, 1, 0.1, and $10^{-2}$.
  The Lyman-$\alpha$ profiles  resulting from perturbation  by either H-He
  or H-H collisions separately are overplotted (dotted lines) for ratios
  of $10^3$ and $10^{-2}$, respectively. The  temperature is 12000~K.}
\label{Hratio}
\end{figure}

\subsection {Collisional profiles simultaneaously perturbed by  He and H atoms}
\label{sec:lyhhhe}

\begin{figure}
\centering
\vspace{8mm}
\includegraphics[width=8cm]{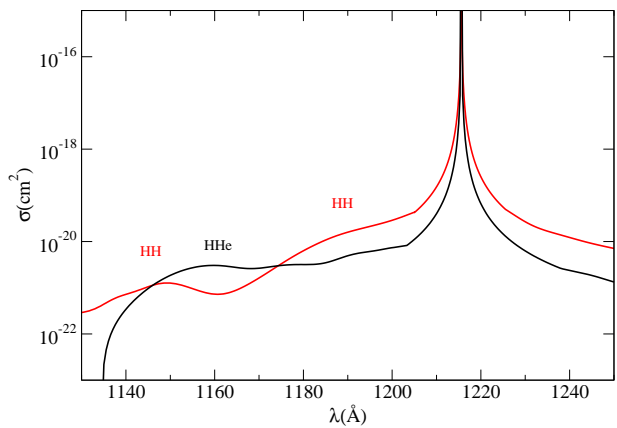}
\caption{Comparison  of the  blue wings of Lyman-$\alpha$ perturbed by
  H-He  collisions (black curve)  with the   blue wing in the 
  Lyman-$\alpha$ range (red curve) due to H-H collisions.
  The contribution of the red wing
  of  Lyman-$\beta$ leads to  a large feature at 1150~\AA\/.
  The H and He densities are 10$^{18}$  cm$^{-3}$, and the  temperature is 12000~K.}
\label{lymanalfabeta18}
\end{figure}

The spectra of helium-dominated white dwarf stars with hydrogen
in their atmosphere present a distinctive broad feature centered
around 1160~\AA\/ in the  blue wing of the  Lyman-$\alpha$ line
(see Fig.~1 in \citet{allard2020}).
Figure~\ref{lymanHH18HHe18} shows that this line satellite is quite  close
to the one due to H-H collisions centered at 1190~\AA\/.
We caused the ratio  $n_{\mathrm{He}}$/$n_{\mathrm{H}}$ of their densities to vary.
Line profiles that are simultaneously perturbed by H and He are  computed for 
a ratio  varying from  $10^{-2}$ to $10^3$, and
the H density $n_{\mathrm{H}}$ remains equal to 10$^{18}$cm$^{-3}$.
When the ratio is $10^3$ , the H-He line profile is
 identical to a pure helium profile, whereas for $10^{-2}$ 
, the H-He line satellite
 at 1160~\AA\/ is not seen.  This is illustrated in Fig.~\ref{Hratio}.
The blue wings of Lyman-$\alpha$
 perturbed by  He or H atoms are compared in Fig.~\ref{lymanalfabeta18}.
The line profile calculations were made at a temperature of 12000~K for 
a perturber density of $10^{18}$~cm$^{-3}$ of He or neutral H.
An additional feature is shown in the blue wing of H-H Lyman-$\alpha$ at 1150~\AA\/.
This feature is a  line satellite of Lyman-$\beta$ quite far from the 
unperturbed Lyman-$\beta$ line center; it is even  
closer to the Lyman-$\alpha$ line. Figure~\ref{lymanalfabeta18} shows that it
is therefore necessary to take the total contribution of the  Lyman-$\alpha$ 
and Lyman-$\beta$ wings of neutral H throughout this region into account.

\begin{figure}
\centering
\vspace{8mm}  
\includegraphics[width=8cm]{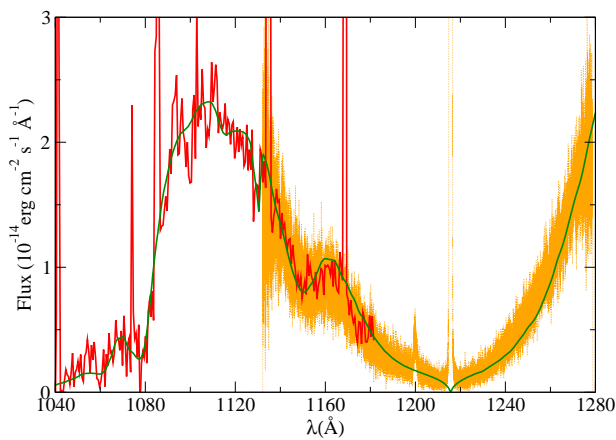}
\caption {Comparison of the  synthetic spectrum obtained with
 $T_{\rm eff} = 12040$~K  and $\log g = 7.93$ (green line)  
  in the Lyman-$\beta$ range with a {\it FUSE}\/
  spectrum of G226-29 (red line)
  (extracted from  Fig.~5 of \citet{allard2004c}).
  We overplot the {\it COS} observation of G226-29
        \citep{hst14076} (orange line).}
\label{G226-29}
\end{figure}

\subsection {Observation of the 1150~\AA\/ satellite}

This  absorption feature due to the 
$B^{\prime\prime}\bar{B}$ $^1\Sigma^+_u$ - $X$ $^1\Sigma^+_g$ transition 
was predicted by \citet{allard2000}.
The ab initio calculations of \citet{spielfiedel2003}
have shown that for the isolated radiating atom
(\mbox{$R \rightarrow \infty$}),
this transition is not asymptotically
forbidden, as was explicitly stated in \citet{allard2000}.

 We reported a  theoretical study of the  variation  of the
 Lyman-$\beta$ profile with the relative density of ionized and neutral atoms
 and demonstrated  that a ratio of 5 of the neutral and proton density is
enough to make this line  satellite  appear in the far wing
(Fig.~5 of  \citet {allard2000}).
The line satellite appearance is then very sensitive to the  degree of
ionization and may be used as a temperature diagnostic.
In \citet{allard2004b}  we reported its first detection
in far-UV (FUV) observations of the pulsating DA white dwarf G226$-$29 obtained with the
{\it Far Ultraviolet Spectroscopic Explorer (FUSE)}. This broad feature was
also detected in the laboratory by \citet{kielkopf2004} and observed in another 
variable DA star, G185-32, by \citet{dupuis2006}.

In \citet{allard2004c} we discussed in detail how important it is  to take the Boltzmann factor in absorption into account in Eq.~(\ref{eq:dip}),
especially in stellar atmospheres for temperatures  lower than 15000~K.
We  considered  local thermal equilibrium model
atmospheres with a pure hydrogen composition that explicitly include the
Lyman-$\alpha$ and Lyman-$\beta$ quasi-molecular opacities.
For the case of G226-29 shown in Fig.~\ref{G226-29} ,
we used a very high signal-to-noise ratio spectrum
obtained using time-resolved
{\it HST}\/ spectra presented by \citet{kepler2000}.
This comparison allowed us to make a temperature and
gravity determination that is compatible with a fit to the {\it FUSE}
observation of this object. 
Fig.~4 of \citet{allard2004b} showed our fit to the {\it HST} spectrum
using our adopted values for $T_{\rm eff} = 12040~K$ and $\log g = 7.93$.
We extracted data from Fig.~5 of \citet{allard2004c}, where 
synthetic spectra in the Lyman-$\beta$ range are compared with  a {\it FUSE}
spectrum of G226-29, to show in
Fig.~\ref{G226-29}  the synthetic spectrum obtained
for  $T$=12040~K and log~$g$=7.93.
 G226-29 was more recently observed with {\it HST~COS}  under program 14076,
and we overplot the  spectrum of
 the G130M$_2$ grating that covers 1130-1270~\AA.
The  {\it HST~COS} observation is noisy but
consistent with the one obtained by  {\it FUSE} and fills  the gap
above 1180~\AA\/ where  the
shoulder in the blue wing of Lyman-$\alpha$ appears.
This part of the spectrum 1180-1200~\AA\/ could not be obtained with 
{\it FUSE} or {\it HST}. However, we should point out that we need to divide the
{\it COS} flux by a factor of $\sim1.3$ to bring it to the  {\it FUSE} spectrum, we had a 
similar problem in \citet{allard2004b} to fit the {\it FUSE}  and {\it IUE}
flux. We considered that this  difference by a factor of $\sim1.3$
in the flux calibration of observations performed with two different
instruments of a faint target was acceptable, but obtaining the same factor
with  {\it COS} would mean that the error is likely due to 
{\it FUSE}.

\section{Conclusions}

The effect of collision broadening by  atomic H and  He
on  spectral lines  is
central for understanding  the opacity of stellar atmospheres.
A correct determination of the Lyman-$\alpha$ line requires the
 determination of  the ground and first excited potential energy curves
 and the electric transition dipole moments with high accuracy. 
 We showed for H-He how important it is to account for relativistic effects to
 characterize  the
 long-distance behavior of the potentials in detail and the avoided
 crossing situations. 
 These effects, although  tiny, become crucial in determining the character and
 the adiabatic  correlation of the states, and in particular, the behavior of
 the associated dipole transition  moments from the ground state, because of the
 specific degeneracy of hydrogen-excited states  in the Coulomb model.  This
 problem does not seem so stringent for H+H ($n$=2) collisions.  Due to
 the $1/R^3$ asymptotic behavior of the states that also undergo
 avoided crossings, but at very large separation, in this case, the behavior
 of the system is expected to be fully diabatic, except perhaps in
ultra-cold and ultra-slow systems.

Our study was conducted assuming
 classical motion for the nuclei, as well as  an adiabatic picture for
 the electronic states during the collisional process.
The so-called diagonal  adiabatic corrections  
\citep{kolos1968,pachucki2014,komasa1993,gherib2016}, which mostly contribute at
short distance, might also be added for an improved accuracy.
Moreover, it should be mentioned that in the regions in which
the potential curves of different states are very close,
such as at large separation  especially in the  H-He case or in short-range
avoided crossing regions such as in H-H, the Born-Oppenheimer approximation
defining the adiabatic states is likely to break down, 
and it might be necessary to take off-diagonal nonadiabatic couplings
and collisional branchings  between the adiabatic states into account.  
\citet{Lique2004} took   the rotational coupling between
states $B^1\Sigma_u^+$  and $C^1\Pi_u$ in H-H collisions into account and  concluded that the
nonadiabatic effects with respect to the adiabatic treatment were very weak
in this case. The effect of nonadiabatic couplings 
remains an open question  in the triplet case, which presents avoided
crossings at short distance.

The {\it HST} observations have motivated this 
  theoretical work, in which we used   accurate  molecular data
  for both the HH \citep{spielfiedel2003,spielfiedel2004}
  and HHe (this work) and extended that of \citet{allard2020}.
  This allowed a thorough study of the atomic underlying atomic physics  
  and accurate  line profile calculations of  Lyman lines
perturbed by collisions with H and He atoms given here. Furthermore,  it
  is also very gratifying to observe features that have been predicted theoretically.
  This was the case of the  1150~\AA\ broad feature in the Lyman-$\beta$ wing
of the {\it FUSE} spectrum  of the DA white dwarf  G226-29 and
now  this one at 1160~\AA\ in the blue wing of {\it COS}  spectra of DBA
white dwarfs.
 The  {\it COS} observation of G226-29 was also 
 an opportunity to reconsider the blue wing of Lyman-$\alpha$.
 Finally, our study is  a  first step
 toward obtaining the accurate data for both Lyman-$\alpha$ and
for Balmer-$\alpha$ that are essential for determining the hydrogen abundance
correctly.


\newpage

\section*{Appendix}

\vspace*{2cm}
\begin{table}

\begin{tabular}{cl}
\hline
&\\
s& 0.03,0.008,0.006902040, 0.0035, 0.00175,0.0008\\
p& 0.12,0.030, 0.015, 0.007,0.0035,0.00175,0.0008\\
d& 0.055406537, 0.024364162, 0.010713761,0.005,0.0025,0.0012\\
f& 0.106396067, 0.046204584, 0.020065249,0.008,0.006,0.0035,0.0015\\
g& 0.168703345, 0.069928301, 0.028985598\\
h& 0.175320015, 0.045069073, 0.011585793\\
&\\
\hline
\end{tabular}
        \caption{ Exponents of Gaussian-type functions on hydrogen added to the aug-cc-pV6Z basis set. The three first exponents of  $d$, $f$, $g,$ and $h$ functions are taken from \citet{allard2020}.}
\end{table}
        \label{tab:basis}

\vspace*{2cm}

\begin{table}
        \centering
        \begin{tabular}{ccccc}
        \hline
                State & ref&  $R_e(\AA)$ &$\omega_e$(cm$^{-1})$ & $D_e$(cm$^{-1}$/eV) \\  
        \hline
                && &\\
                $A\Sigma$& (a) & 0.74022 & 3730.3  & 20658 (2.561)\\
                           &(b)& 0.74074& 3697.2  &  (2.563\\
                &(c)& 0.7472& 3726& (2.54)\\
                &(d)& 0.7430& 3512\\ 
                &(e)& 0.7573& 3662 \\
                &(f)& 0.7567& 3701\\ 
                &(g)& 0.74086& 3718 \\
                && &\\
                $B\Pi$& (a)& 0.76813 & 3339.9 &  17868.4 (2.218) \\
                      & (b)&0.76863& 3313.4 & (2.218)\\
                &(c)& 0.7738& 3372&  (2.20)\\
                &(d)& 0.7711& 3158\\
                &(e)& 0.7741& 3302& (2.20)\\
                && &\\
                $C\Sigma$ & (a)&  0.80927 & 2916.5 & 13665.5 (1.694) \\
                          & (b)&  0.80953& 2906.3&(1.638)\\
                &(c)& 0.8133& 2957& (1.61)\\
                &(d)& 0.8073& 2788\\
                &(e)& 0.81641& 2872& (1.65)\\
                &(f)& 0.8308& 2896 \\
                &(g)& 0.8109& 2902\\

        \hline
        \end{tabular}
        \caption{Spectroscopic constants of the molecular states dissociating into  H(2s,2p)+He(1s$^2$).
        $D_e$ values in parentheses are in  eV.  The dissocation energy of state $C\Sigma$  is taken below the barrier top lying 5386 cm$^{-1}$ above the dissociation limit  H(2s,2p)+He.  (a) this work, (b) theory, \citet{allard2020}, (c) theory, \citet{lo2006}, (d) theory, \citet{sarpal1991}, (e) theory, \citet{theodora1987}, (f) experimental spectroscopy, \citet{ketterle85}, and (g) experimental spectroscopy, \citet{ketterle88}.}
        \label{tab:e2}
\end{table}

\end{document}